\documentclass{ceurart}

\usepackage{listings}
\begin{document}

\copyrightyear{2025}
\copyrightclause{Copyright for this paper by its authors.
  Use permitted under Creative Commons License Attribution 4.0
  International (CC BY 4.0).}

\conference{AI4A2J-ICAIL25: AI for Access to Justice at the International Conference on AI and Law 2025,
  June 20, 2025, Chicago, IL}

\title{Place Matters: Comparing LLM Hallucination Rates for Place-Based Legal Queries}


\author[1,2]{Damian Curran}[%
orcid=0009-0006-1440-311X,
email=curran.d@unimelb.edu.au,
]
\cormark[1]

\author[2]{Vanessa Sporne}

\author[1]{Dr. Lea Frermann}[%
orcid=0000-0002-9712-1188,
email=lea.frermann@unimelb.edu.au,
]

\author[2,3]{Prof. Jeannie Paterson}[%
orcid=0000-0002-2649-7591,
email=jeanniep@unimelb.edu.au,
]

\address[1]{School of Computing and Information Systems, The University of Melbourne, Australia}
\address[2]{The Centre for Artificial Intelligence and Digital Ethics}
\address[3]{Melbourne Law School, The University of Melbourne, Australia}

\cortext[1]{Corresponding author.}


\begin{abstract}
    How do we make a meaningful comparison of a large language model's knowledge of the law in one place compared to another? Quantifying these differences is critical to understanding if the quality of the legal information obtained by users of LLM-based chatbots varies depending on their location. However, obtaining meaningful comparative metrics is challenging because legal institutions in different places are not themselves easily comparable. In this work we propose a methodology to obtain place-to-place metrics based on the comparative law concept of functionalism. We construct a dataset of factual scenarios drawn from Reddit posts by users seeking legal advice for family, housing, employment, crime and traffic issues. We use these to elicit a summary of a law from the LLM relevant to each scenario in Los Angeles, London and Sydney. These summaries, typically of a legislative provision, are manually evaluated for hallucinations. We show that the rate of hallucination of legal information by leading closed-source LLMs is significantly associated with place. This suggests that the quality of legal solutions provided by these models is not evenly distributed across geography. Additionally, we show a strong negative correlation between hallucination rate and the frequency of the majority response when the LLM is sampled multiple times, suggesting a measure of uncertainty of model predictions of legal facts.
\end{abstract}


\maketitle


\section{Introduction}
How do we make a meaningful comparison of a large language model's (\textbf{LLM}) knowledge of the law in one place compared to another? For instance, does an LLM have 'better' knowledge of the laws of Los Angeles, USA, than the laws of London, UK? Will a person using an LLM chatbot to assist with a legal problem in Sydney, Australia, be provided with legal information that is equally as accurate as a person in a similar situation, but subject to the laws of Toronto, Canada? Or Singapore? Or Auckland? 

These questions are important because a growing body of survey work suggests that ordinary people across the globe (including Czechia \cite{kuk_llms_2025}, the UK \cite{seabrooke_survey_2024} and California \cite{hagan_towards_2024}) are increasingly using LLM-based chatbots to obtain legal information for personal matters. The ostensible reason for the suitability of LLM chatbots for these tasks is obvious - LLM access is either free or relatively low cost, has an accessible chat interface and can provide personalized information, targeted to individual situations. But, as noted by Dahl et al., the potential for LLMs to close the access to justice gap will only be reached if LLMs 'actually know the law' \cite{dahl_large_2024}. If an LLM provides inaccurate legal information which nonetheless sounds plausible, it may in fact have the opposite effect on the justice gap by systematically disadvantaging those who need the most help. 

In this work we explore whether the accuracy of legal information produced by LLMs varies for different users across geographic boundaries. It is a trite observation - but one worth noting explicitly - that laws vary from place to place. Whilst some high-level norms may be globally consistent - such as the near universal prohibition on violent crime - the actual form and manifestation of those norms will vary, often to a local province or county level. Very little of the legal landscape is truly international and those few subject areas that are international have very little bearing on the typical pro se (or 'self-represented') litigant seeking legal advice for a personal matter, such as a family, employment or tenancy dispute. Most such laws vary along national and often local boundaries and so an LLM's knowledge of such a law in one place will not necesarily be indicative of comparable knowledge in another. Understanding how hallucination rates vary across geographic bounds is therefore critical for evaluating whether access to justice can be delivered equitably to all LLM users, regardless of where they live.

In this work we propose a new methodology for obtaining a metric of LLM hallucination rates which can be used to meaningfully compare LLM legal knowledge across geographic lines. We discuss the method's theory, drawing on the comparative law concept of functionalism. We provide empirical evidence that the rate of hallucination of legal facts by leading closed-source LLMs is significantly associated with place and we begin a discussion on possible correlating factors based on observations in our data.

\begin{figure}
  \centering
  \includegraphics[width=\linewidth]{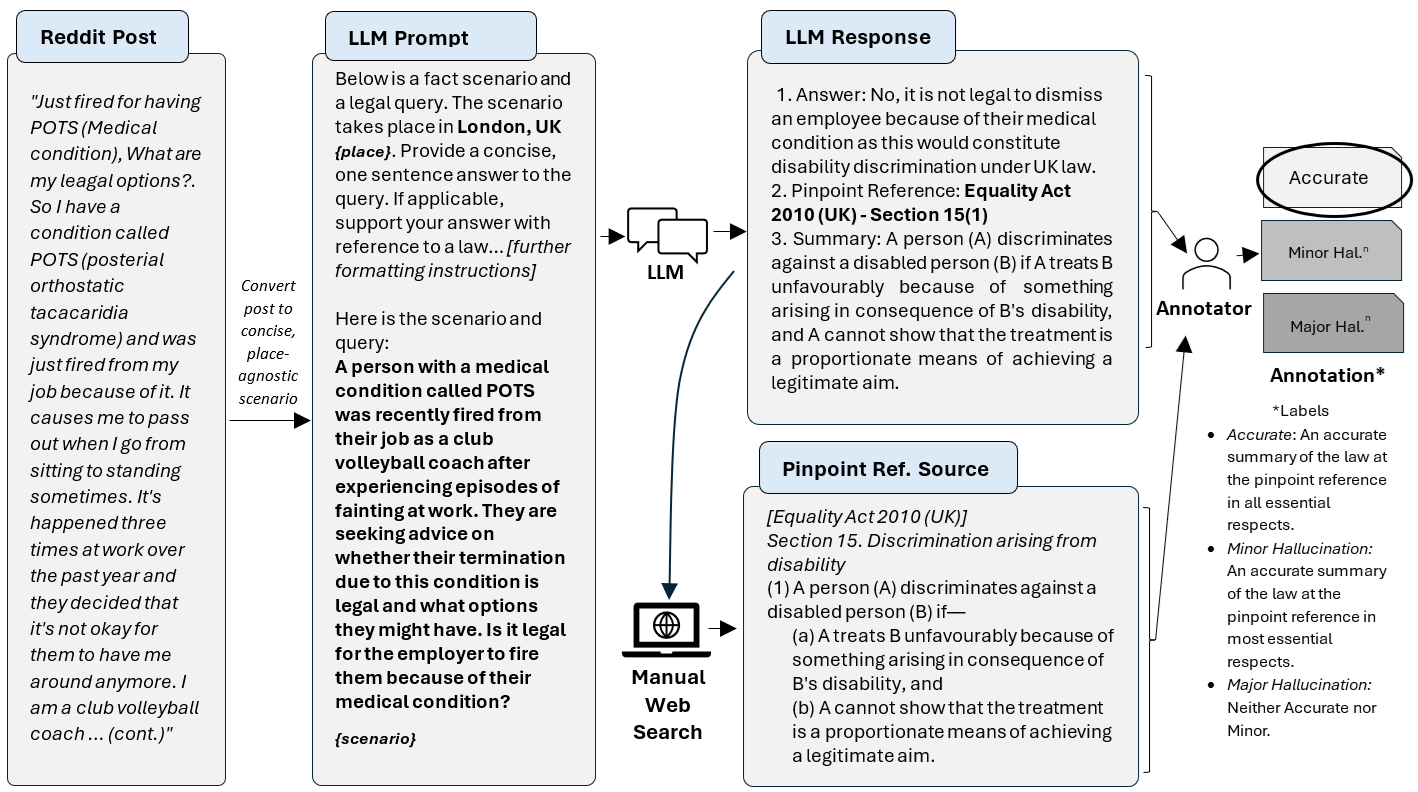}
  \caption{\textbf{Overall workflow}: A subject matter-diverse set of Reddit posts were converted to concise, place-agnostic scenerios. The prompt template has slots for the 'place' and the converted 'scenario'. The source text of the pinpoint reference identified by the LLM is found online, and the LLM's summary is manually annotated for hallucinations by a human annotator.}
  \label{fig:workflow}
\end{figure}

\section{A Comparative Law Approach}

Producing meaningful comparative metrics of an LLM's knowledge of laws in different places is challenging. One method that might initially be considered is a comparison of the hallucination rates on datasets of conceptually similar legal instruments in different places, such as the body of laws passed by the legislatures in the places being examined. But whilst these metrics would be useful for evaluating and improving the localized performance of the models in each place, there are significant challenges in using them as the basis for an 'apples-to-apples' comparison of LLM knowledge across places. This is because each body of legislation has a different size and subject matter. They also operate in different historical, political and cultural contexts.

Thankfully, this is not a new conceptual problem. Comparative lawyers have thinking about how to compare different legal systems for generations. One method in particular has come to dominate modern comparative law studies – namely, functionalism. Functionalism can be defined as having at its core 'the comparison of solutions which different legal orders offer for specific practical problems' \cite{kischel_comparative_2019}. Instead of attempting to compare conceptually similar legal institutions and instruments, we instead begin with specific practical problems, and compare the solutions offered to those problems in each place. We adopt a functionalist approach in this study by framing our dataset as a series of 'practical problems' to which the LLM needs to identify a solution in each place. The 'solution', typically a reference to a legal instrument applicable in the place, is then evaluated for hallucinations. The resultant hallucination rates can then be meaningfully compared, because they reflect the LLMs relative knowledge of the solutions to the \textit{same} practical problems.

\section{Related Work}

\textit{Hallucination Definitions and Prevalence in Legal Tasks.} There is no generally accepted definition of an LLM hallucination \cite{van_deemter_pitfalls_2024,ji_survey_2023}. Common classification schema differentiate between whether the hallucination is inconsistent with the training data or the input prompt (such as the 'open-' or 'closed-' domain described by Bubeck et al. \cite{bubeck_sparks_2023}), or whether it is non-factual or unfaithful to the user instructions (\cite{huang_survey_2025}). In the context of hallucination of legal facts, Dahl et al. \cite{dahl_large_2024} defined hallucination as 'a factual infidelity between an LLM's response and the controlling legal landscape.' They found that public-facing LLM chatbots frequently hallucinate legal facts. Magesh et al. \cite{magesh_hallucination-free_2024} expanded on this work finding that even retrieval-based legal research tools produce hallucinated outputs.

\textit{Geographic bias.} Several studies have shown that performance of LLMs varies along geographic dimensions. Moayeri et al. developed a dataset by which the same question, such as population level, could be asked about different countries \cite{moayeri_worldbench_2024}. They observed consistent disparities in LLM knowledge between countries in Sub-Saharan Africa compared to North America. Manvi et al. showed that LLMs exhibit biases against certain locations with lower socio-economic conditions across a range of objective and subjective topics \cite{manvi_large_2024}. Dahl et al. identified discrepancies between LLM knowledge of caselaw originating from different U.S. federal court circuits, which are geographically defined \cite{dahl_large_2024}. 

\textit{Self-consistency.} One promising method of automatically identifying and reducing hallucinated content from LLMs is self-consistency. Self-consistency methods sample multiple responses and select the most consistent answer via majority voting \cite{cheng_integrative_2024,wang_fine-grained_2024}. This method has been shown to improve language model performance on a range of arithmetic and common-sense benchmarks [20]. Brown et al. showed that language model performance in a number of domains improves as the number of samples produced at inference time increases \cite{brown_large_2024}.

\section{Method}

\subsection{Workflow \& Dataset}

In each instance we prompt the LLMs with a legal scenario said to occur in a particular place. We ask the models to identify and summarise an applicable law. Legally qualified annotators then manually locate that law online and evaluate whether the LLM output is an accurate summary of the actual law. Summaries that are not accurate are labelled either minor or major hallucinations. We then analyze the resulting hallucination rates and their association with the place. The full workflow is illustrated in Figure \ref{fig:workflow}.

Each instance of our dataset is comprised of a concise factual scenario and a query about a related legal issue. The scenarios originate from a sample of online posts from the \textit{'r/legaladvice'} sub-Reddit, which in turn were compiled by the Learned Hands Project \cite{suffolk_lit_lab_and_stanford_legal_design_lab_learned_2018}. We have used this dataset because the narratives reflect actual queries about real legal problems encountered by individuals, who then seek advice online.\footnote{We use Reddit posts as a proxy for the types of queries people might ask a chatbot. We are not aware of any user logs from the closed-source chatbots which are available for research purposes.} Our goal was to buidl a dataset with a high diversity of legal topics and sub-topics. We initially labelled a large collection of instances using the lower tiers of the LIST Legal Issues Taxonomy classification scheme \cite{the_leland_stanford_junior_university_stanford_university_list_2024}. We then manually selected combinations of scenarios which provided high label diversity within four high-level legal issues. Our final 100 instances comprise 25 examples each from the posts originally labelled with 'tenancy \& housing' (\textbf{Housing}), 'employment law' (\textbf{Employment}) and 'family and custody matters' (\textbf{Family}), and 25 examples in a combined category of 'crime' and 'traffic' (\textbf{Crime \& Traffic}) (each category a \textbf{legal issue}). 

In order to ensure a consistent format, we re-drafted the selected posts into a concise factual scenario of one or two sentences, plus a single sentence query based upon the user's questions. We standardized the tone to a third person, objective voice, and anonymized the text. Because these queries are to be used to compare the laws in different places, we also removed any terminology which obviously identified the query as originating in one particular place (e.g. the names of particular courts, or bureaucratic institutions such as 'DMV'). We referred to this process as making the scenarios 'place-agnostic'. We end up with a dataset of 100 diverse, place-agnostic scenarios across four broad legal topics. An example of an original Reddit post and the converted scenario is shown in Figure \ref{fig:workflow}.

\begin{figure}
  \centering
  \includegraphics[width=\linewidth]{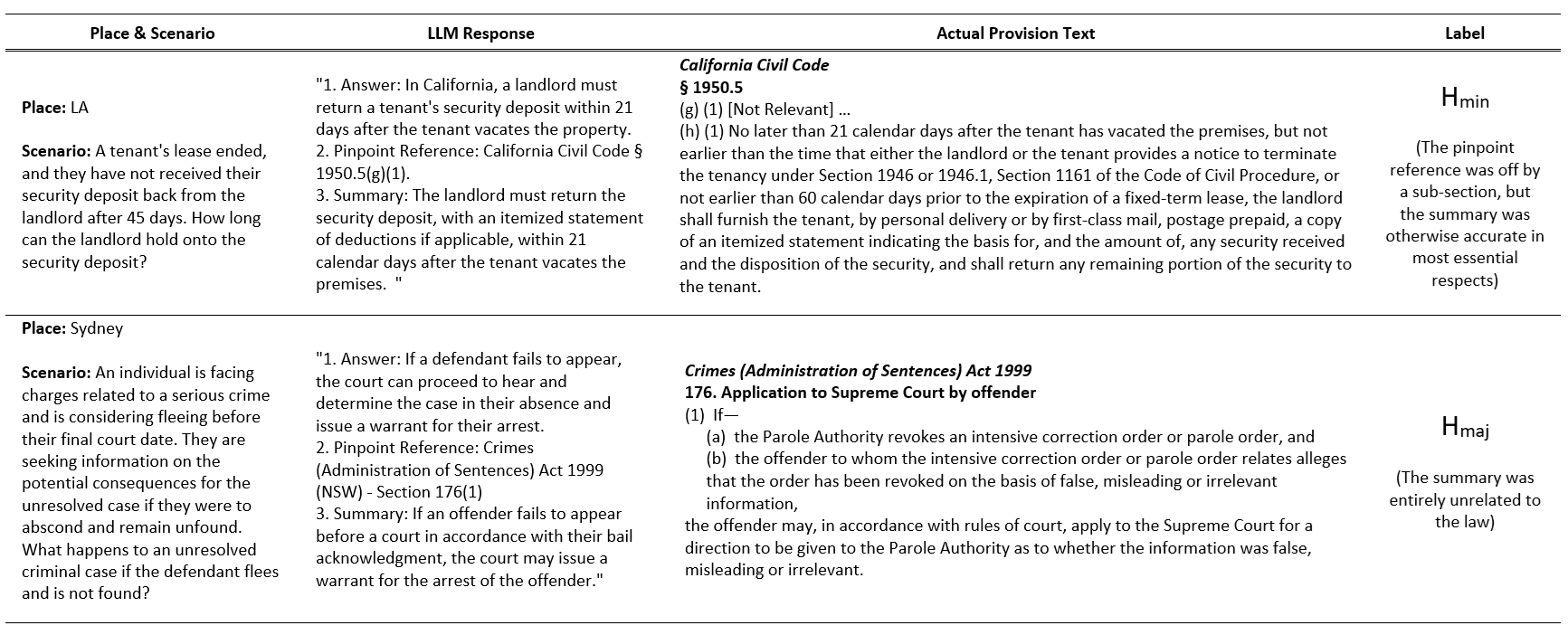}
  \caption{\textbf{Annotations}: Examples of our $\mathbf{H_{min}}$ and $\mathbf{H_{maj}}$ annotations. An example \textbf{A} annotation can be found in Figure \ref{fig:workflow}}
  \label{fig:example_annots}
\end{figure}

\subsection{Places}

We test our approach on three places, namely \textit{'Los Angeles, USA'} (\textbf{LA}), \textit{'London, UK'} (\textbf{London}), and \textit{'Sydney, Australia'} (\textbf{Sydney}). We select large population centers from different countries where the laws are written in English. We suspect that differences may be more more pronounced across other places operating in lower-resource languages, but wished to reduce the number confounding factors in this initial study. We specify exact locations to the city-level, so that any scenario which happens to be resolved by local or council ordinances and rules, such as planning law and traffic violations, can be addressed with specificity.

\subsection{Prompt \& Response}

The prompt asks the LLM to answer the legal query and then, if applicable, identify a law that supports that answer. We use a prompt template with slots for both the scenario and the place. The LLM output should include an answer (\textbf{LLM Answer}), a law with pinpoint reference (\textbf{LLM Reference}), and a summary of the law at that reference (\textbf{LLM Summary}). Whilst we were not evaluating the correctness of the LLM Answer directly, our view was that asking for an answer in the prompt was a closer reflection of how a pro se (or 'self-represented') litigant might naturally use an LLM chatbot to seek information and that we ought to evaluate outputs which were conditioned on the presence of the LLM Answer.

\subsection{Models \& Sampling}

We test the production LLMs which underpin popular, consumer chatbots from OpenAI, Anthropic and Google at the time of testing (late-2024), namely GPT-4o (\textit{gpt-4o-2024-11-20}) (\textbf{GPT-4}), Claude-Sonnet-3.5 (\textit{claude-3-5-sonnet-20241022}) (\textbf{Claude}) and Gemini-Pro-1.5 (\textit{gemini-1.5-pro-002}) (\textbf{Gemini}). We access the models via their API, with temperature set to 0.0.\footnote{For reproducibility, we will make available our code, dataset, prompts, LLM responses, annotation guide and annotations at our code repository.}

We found in early testing that the models often produce variations in their responses to identical queries. For every scenario-place-LLM combination, we therefore sample the model 10 times. We use majority voting to extract the response which was produced most frequently in the 10 samples using exact, verbatim string matching. If there was a tie, we selected the first instance alphabetically. We evaluate only that response.\footnote{In a small number of cases involving the Google Gemini model (n=8), the majority sample was a refusal to produce any response. In those instances, the second most frequent response for that scenario was evaluated.} We record the frequency of occurrence, out of 10, of the majority samples (\textbf{majority sample frequency}) (i.e. a majority sample frequency of 1 indicates that every response in the 10 samples was different, whereas a majority sample frequency of 10 indicates that every response in the 10 samples was identical.) We do this for two reasons. First, it increases robustness of our hallucination metrics, making them less prone to random chance variations in the responses. Second, it allows us to contribute to the literature on automated hallucination detection. Other studies have shown that responses that consistently repeat over multiple samples of LLMs are less likely to be hallucinated \cite{brown_large_2024}. We report on correlations between hallucination rate and the majority sample frequencies using Spearman's rank correlation coefficient.

\subsection{Annotation \& Metrics}

We evaluated whether the LLM outputs were consistent with real-world legal facts - i.e. what \cite{huang_survey_2025} called 'factuality' hallucinations, or alternatively what \cite{dahl_large_2024} defined as 'a factual infidelity between an LLM's response and the controlling legal landscape.' Two of the authors, who are both qualified lawyers with practicing experience, manually annotated the output by locating the LLM Reference online (if it exists) and comparing it to the LLM Summary. We apply one of these four labels to each instance:

\textit{1. Accurate} (\textbf{A}): The LLM Summary is considered Accurate if it is an accurate summary of the law at the LLM Reference in \textit{all} essential respects. 

\textit{2. Minor Hallucination} ($\mathbf{H_{min}}$): The LLM Summary is a Minor Hallucination if it is an accurate summary of the law at the LLM Reference in \textit{most} essential respects. We use this category to identify cases where the output is not Accurate, but we wish to recognize that the LLM nonetheless has decent knowledge of the law in this area. A Minor Hallucination can include where the pinpoint reference was slightly off (e.g. a reference to \textit{s.5(a)(i)} instead of the correct \textit{s.5(a)(ii)}), or where the summary contains some additional information not found at the pinpoint, but which is likely to be correct given the context and does not change the substantive meaning of the balance of the LLM Summary.

\textit{3. Major Hallucination} ($\mathbf{H_{maj}}$): The LLM Summary is neither Accurate nor a Minor Hallucination.

\textit{4. No Law} (\textbf{NL}): We appreciate that there may be no written laws that apply to a particular scenario in a particular place. In the prompt we invite the LLM to identify those instances. Such responses are annotated with a No Law label.

We prepared an annotation guide to the above effect, with examples, prior to annotating. Each annotator received an equal portion of instances stratified along the place, legal issue and model dimensions. A total of 216 instances, or 24\% of the dataset, stratified along place, model and legal issue dimensions, were annotated by each annotator to establish inter-annotator agreement. Examples of the nature of the LLM outputs and our annotations are shown in Figure \ref{fig:example_annots}. 

We report two hallucination rate measures. The hallucination rate, \textit{hr}, is the average rate at which any hallucination is produced in the population. We also report on the major hallucination rate, \textit{hr*}, measuring only the occurrence of major hallucinations. These equations are shown in below. Given the place and the annotation label are categorical, we perform a Chi-Square test of independence to identify if there is a significant relationship between place and hallucination rates.

\[
\textit{hr}  = \frac{\text{count}(\mathbf{H}_{\text{maj}}, \mathbf{H}_{\text{min}})}{\text{count}(\mathbf{H}_{\text{maj}}, \mathbf{H}_{\text{min}}, A, NL)}, \quad
\textit{hr*} = \frac{\text{count}(\mathbf{H}_{\text{maj}})}{\text{count}(\mathbf{H}_{\text{maj}}, \mathbf{H}_{\text{min}}, A, NL)}
\]

\section{Results}

\subsection{Annotator Agreement}

We evaluated 900 instances split equally across three models and three places. Excluding 15 NL instances, there were 201 instances labelled by both annotators (~25\% of the total non-NL labelled data). Cohen's kappa coefficient of inter-annotator agreement on these queries was $\kappa$ = 0.66. A score of over 0.60 is generally considered 'substantial' agreement between annotators \cite{landis_measurement_1977}. This suggests the category labels and instructions were able to guide annotators to common labels in most situations.

\subsection{Hallucination Rates and Association to Place}

\begin{figure}
  \centering
  \includegraphics[width=\linewidth]{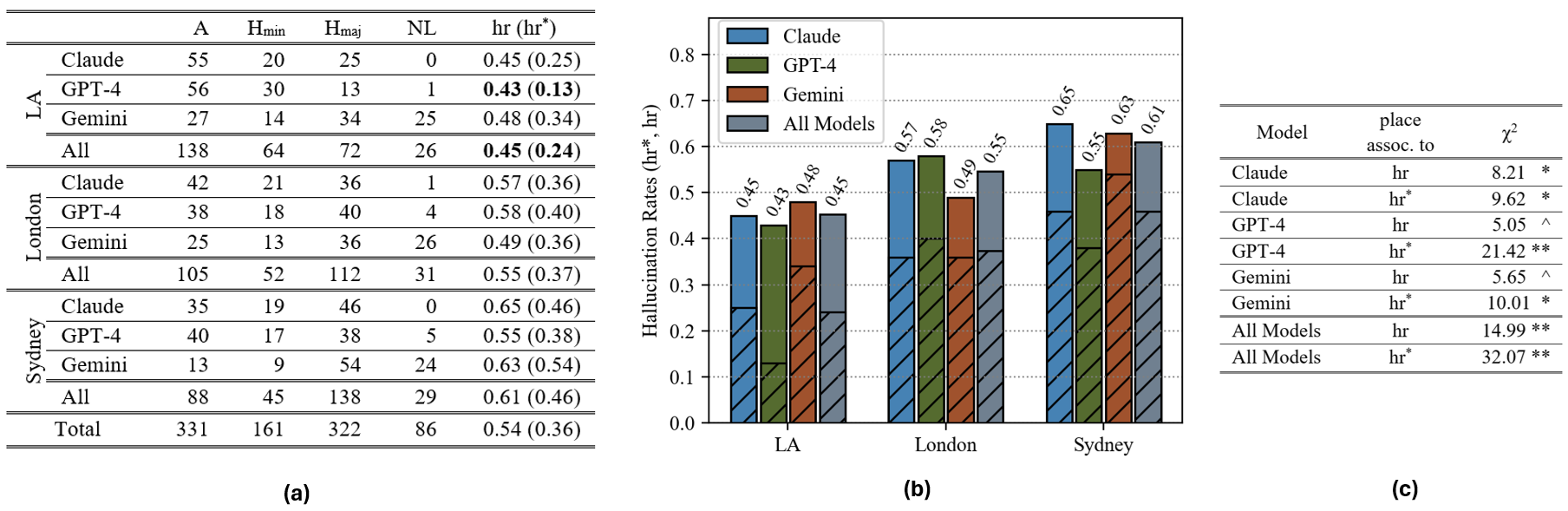}
  \caption{\textbf{Association of Hallucination Rates to Place}: Figure(a): Hallucination rates by model. Figure(b): Rates by model, plotted. Figure(c): Chi-Square statistic for each association to place. Significance of results marked ** ($p < 0.001$), * ($p < 0.05$), or $\text{\textasciicircum}$ ($p < 0.1$).}
  \label{fig:hall_rates-place}
\end{figure}

The label counts and resulting hallucination rates, by model and place, are shown in Figure \ref{fig:hall_rates-place}(a). The hallucination rates are plotted in Figure \ref{fig:hall_rates-place}(b). Across all models, LA had the lowest hallucination rates (\textit{hr}=0.45), followed by London (\textit{hr}=0.55) and then Sydney (\textit{hr}=0.61). The best performing model was GPT-4 in LA, with \textit{hr} of 0.43. and \textit{hr*} of 0.13. The highest \textit{hr} was Claude in Sydney with 0.65. Across all places, GPT-4 is the best performing model, followed by Gemini and Claude (with \textit{hr}'s respectively of 0.52, 0.53 and 0.56.) 

The chi-square test shows strong and significant association between place and \textit{hr} ($\chi^2$=14.99) and between place and \textit{hr*} ($\chi^2$=32.07). This indicates that the likelihood of hallucination of legal facts (and LLM knowledge of legal facts) is not evenly distributed and varies depending on place. All chi-square statistics are shown in Figure \ref{fig:hall_rates-place}(c).

We were cognizant of the risk that by annotating only the LLM Summary and LLM Reference (but not its answer to the legal query itself), the resulting hallucination rates may be disconnected from the original legal query. To gauge the extent to which this may have occurred, we carried out two additional checks for every evaluated instance. We first confirm that the LLM Answer responded to the query being asked, and then confirm that the LLM Summary appeared to support the whole or part of the LLM Answer. We identified only 22 of 900 instances (<2.5\%) in which one or both checks could not be confirmed. These few instances were relatively evenly distributed across place, legal issue and model. This suggests that the LLM outputs are internally consistent between the prompt query, the answer and the law summary, and hence that our hallucination rates are a good reflection of the LLM knowledge of the law related to the scenario in each place.

\subsection{Correlation of Sample Frequency and Hallucination Rates}

\begin{figure}
  \centering
  \includegraphics[width=\linewidth]{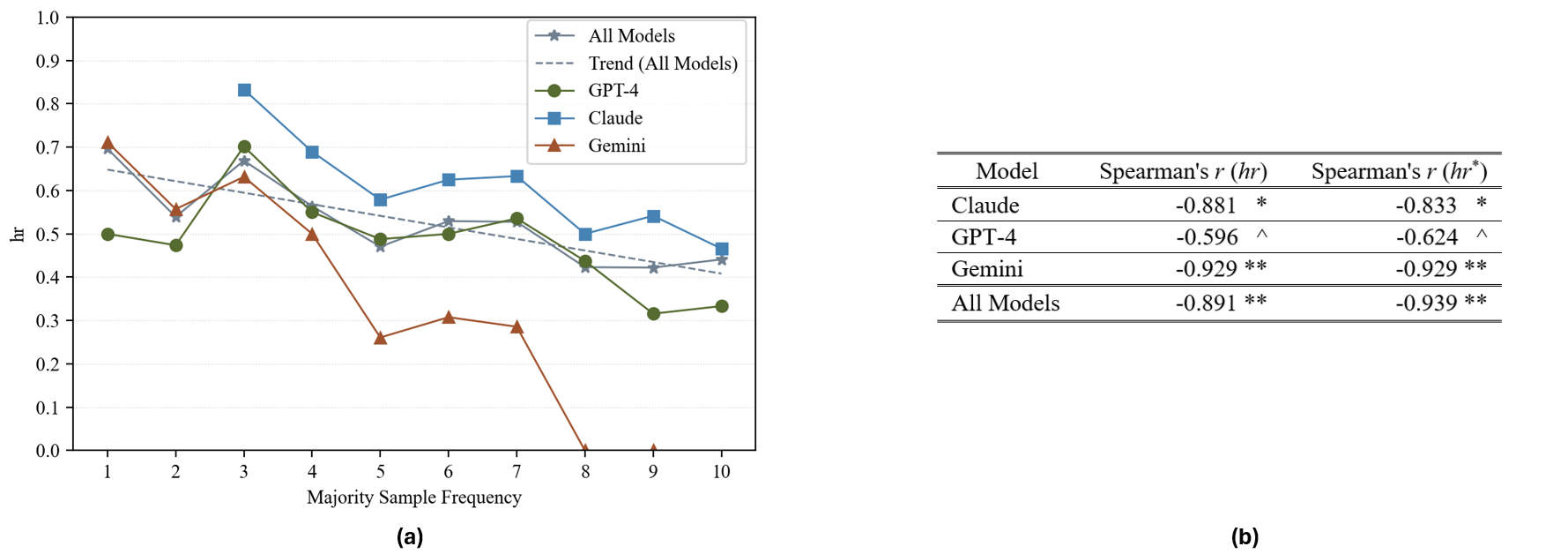}
  \caption{\textbf{Correlation of Sample Frequency and hr}: Figure (a): \textit{hr} and majority sample frequency, by model. Figure (b): Spearman's rank correlation coefficients between majority sample frequency and \textit{hr} and \textit{hr*}. ** ($p < 0.001$), * ($p < 0.05$), or $\text{\textasciicircum}$ ($p < 0.1$).}
  \label{fig:sample_freq}
\end{figure}

We conducted a Spearman's rank correlation analysis to identify whether there was a correlation between majority sample frequency and the hallucination rates, \textit{hr} and \textit{hr*}. A Spearman coefficient of -1.0 indicates a perfect negative association. There is a strong negative correlation between majority sample frequency and \textit{hr} and \textit{hr*} (r=-0.891 (\textit{hr}) and r=-0.939 (\textit{hr*}).) This suggests that an increase of the majority sample frequency is associated with fewer hallucinations of legal facts. The \textit{hr} for each frequency and model is shown in Figure \ref{fig:sample_freq}(a), along with a linear trendline for all models. The decrease in the \textit{hr} is evident in the trend line as majority sample frequency increases. The rank coefficients are shown in Figure \ref{fig:sample_freq}(b).

\subsection{Nature of LLM References}

We conducted a series of analyses on the nature of the laws identified in the LLM References in each place. The vast majority of laws (88.5\%) in the LLM References were codified instruments such as statute, code, regulations or rules. 9.5\% of instances were No Law. Only 1.4\% were case law and 0.4\% were references to a constitutional provision. There was moderate agreement across the models as to the most relevant law in each place for each scenario. Specifically, all three models identified the same law in the LLM Reference for each scenario-place combination approximately 40\% of the time. However, agreement across the models as to the most relevant pinpoint reference was less common, occurring only 8\% of the time. Responses in which the law in the LLM Reference did not exist at all were very rare (0.5\%). The balance of hallucinations were either a non-existent pinpoint reference within the instrument or a hallucinated LLM Summary.

\begin{figure}
  \centering
  \includegraphics[width=\linewidth]{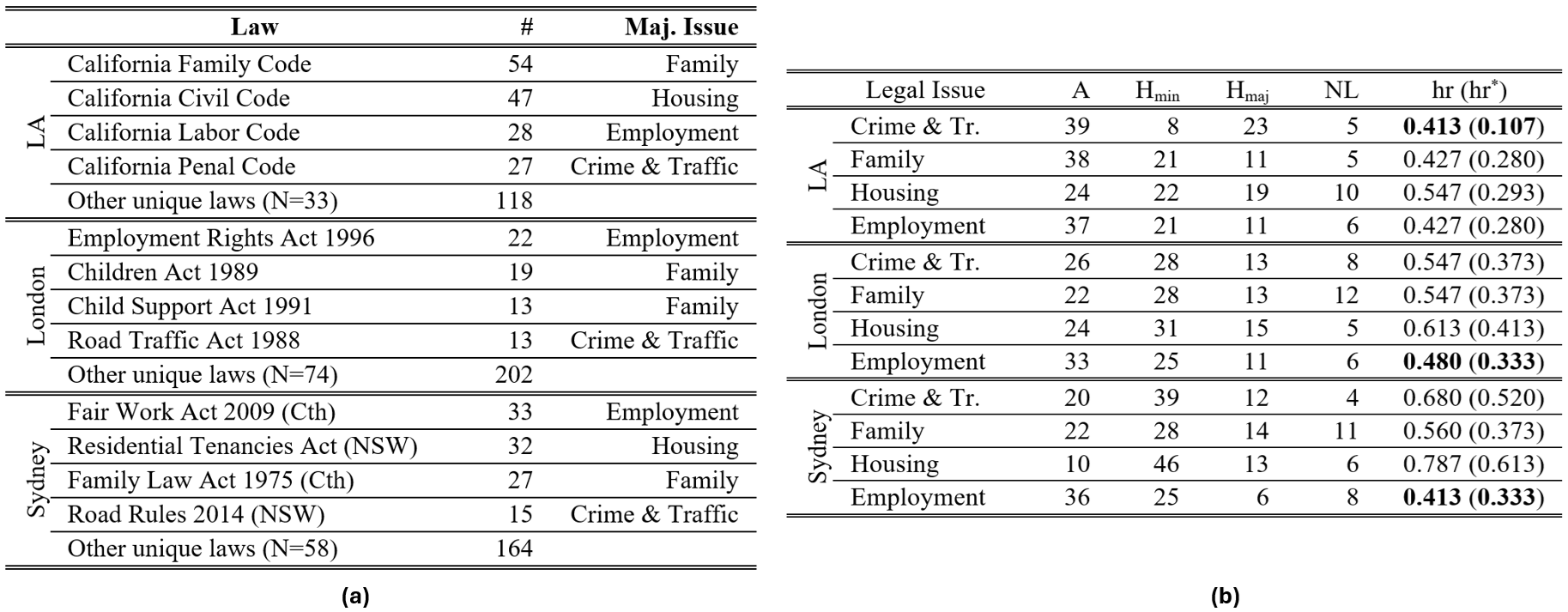}
  \caption{\textbf{Nature of LLM References}: Figure(a): Most frequently cited laws in each place across all models, number of instances and majority legal issue. Figure(b): Hallucination rates and counts by legal issue.}
  \label{fig:other_analysis}
\end{figure}

We measured the diversity of the laws cited in the LLM References across each place. LA had the lowest number of unique laws cited, with 37 unique laws, then Sydney with 62 and London with 78. The four most frequently cited laws and the tail length in each place across all models is shown in Figure \ref{fig:other_analysis}(a). LA has the shortest tail of unique laws. It had a much greater concentration of mentions in its most frequently mentioned laws, specifically the California Family, Civil, Labor and Penal codes. This is notable compared to London, where over twice as many unique laws were identified than for LA. The family law queries in LA, for example, often referenced a single law – the California Family Code – as opposed to the Family law queries in London, which were addressed with reference to several instruments including the Children Act 1989, the Child Support Act 1991 and the Matrimonial Causes Act 1973. This suggests that the laws in LA (or at least in the State of California) might be more densely organized into fewer instruments than in the other places.

We also analyzed the divergent hallucination rates across legal issue. Crime \& Traffic in LA has the lowest rates for any place-legal issue combination. Across all places, Housing had the highest hallucination rates with an average \textit{hr} of 0.649, followed by Crime \& Traffic (\textit{hr}=0.547), Family (\textit{hr}=0.511) and Employment (\textit{hr} = 0.440). The counts and rates by legal issue and place are shown in Figure \ref{fig:other_analysis}(b).

\section{Discussion and Future Work}

It will come as no surprise to readers that LLMs hallucinate facts about the law. What this work contributes is evidence that there are systematic differences in the rate of those hallucinations across place. Pro se litigants should be aware that the quality of the information they may receive from chatbots powered by these models may vary depending on where they live. Legal service providers that incorporate LLM outputs into their workflows should be aware that evaluation of LLM performance in one jurisdiction may not translate to another.

Whilst this study has shown that there is an association between hallucination rates and place, we have not sought to establish cause or correlations. However, we make some suggestions here for future research in this area. Technical explanations for the association may relate to the training data distributions. Research from Kandpal et al. suggests that LLMs exhibit increased hallucination rates when they generate text relating to entities which appear less frequently in the training datasets \cite{kandpal_large_2023}. They refer to this as an LLM's struggle to learn 'long-tail knowledge'. So, if the training data includes more mentions about a particular law - perhaps manifested as a greater number of relevant online legal texts, law firm websites or blog posts - then the hallucination rate on queries about that law may decrease. Whilst we do not have access to the training data of the closed-source models we evaluated to confirm this intuition, it is plausible that the training data comprised of web-content from a larger jurisdiction could contain many more references to their laws than those from smaller jurisdictions. That may partly explain the relatively lower \textit{hr} in Los Angeles (as part of the State of California and its 39m. people) compared to, say, Sydney (as part of the State of NSW and its 8m. people). 

This intuition is further supported by closer analysis of the Syndey data. Take the two laws most-cited by the LLMs in Sydney, the Fair Work Act 2009 (FWA) and the Residential Tenancies Act 2010 (RTA), cited 33 and 32 times respectively (see Figure \ref{fig:other_analysis}(a)). The FWA is federal legislation, which applies across the entire country of Australia. LLM responses which mentioned the FWA had a combined hallucination rate of \textit{hr}=0.394. The RTA, on the other hand, only applies at the state level. LLM responses which mentioned the RTA had a 100\% \textit{hr} (i.e. the LLMs \textit{never} got it right!). Additionally, the family law matters (predominantly regulated by federal legislation in Australia) had a \textit{hr} of 0.560, whilst crime and traffic matters (predominantly regulated at the state level in NSW) had a notably higher \textit{hr} of 0.680. This suggests that the rate of hallucination of legal facts in much smaller jursidictions may be even higher.

We also find that increasing the number of times an LLM is sampled can produce signals which indicate the hallucination of legal facts. Majority sampling could be utilized as a measure of uncertainty of predictions of legal facts. Whilst taking multiple samples incurs greater computational cost, that may be an acceptable trade-off given the low-volume, high-importance nature of many legal queries.

Future work could tease out the impact of place across a greater diversity of legal issues and languages. This work only examined three relatively populous places in English - the world's top resourced language. If there is indeed a correlation between the size of the jurisdiction and the hallucination rates, the rates are likely to be even higher in the long tail of small and medium sized jurisdictions using lower resource languages.

\section{Conclusion}

In this work we proposed a new methodology for quantifying differences in a LLM's knowledge of the law in one place compared to another. Quantifying these differences is critical to understanding if the quality of the legal information obtained by users of these tools varies depending on their location. We discussed why obtaining meaningful comparative metrics is challenging because legal instruments in different places are not themselves easily comparable. We then proposed a methodology based on the comparative law concept of functionalism. We constructed a dataset of legal 'problems' to elicit a set of 'solutions' from the LLM relevant to each place. We ran our dataset over three places (Los Angeles, London and Sydney) and three LLMs (GPT-4, Claude and Gemini). The LLM outputs were manually evaluated for hallucinations. Our analysis showed that the rate of hallucination of legal facts is significantly associated with place. We discussed whether hallucination rates may be correlated to the training data distribution using jurisdiction size as a proxy. We also contributed to the literature on automated hallucination detection by showing a strong negative correlation between hallucination rate and the size of the majority-sample in the context of generated legal facts.  

\section{Declaration on Generative AI}

The authors have not employed any Generative AI tools.

\begin{acknowledgments}

This work was made possible due to generous support of the Centre for Artificial Intelligence and Digital Ethics (CAIDE) at the University of Melbourne.

\end{acknowledgments}

\bibliographystyle{plainnat} 
\bibliography{PlaceMatters}

\end{document}